\documentclass[preprint2]{aastex62}
\usepackage{textcomp}
\usepackage{gensymb}
\usepackage{subfigure}
\usepackage{amsmath}

\shorttitle{Correction to the photometric magnitudes of the {\it Gaia} EDR3}
\shortauthors{Yang et al.}

\DeclareGraphicsExtensions{.eps,.pdf,.jpg,.bmp}

\begin{document}
\title{Correction to the photometric magnitudes of the {\it Gaia} Early Data Release 3}
\author{Lin Yang}
\affil{College of Artificial Intelligence, Beijing Normal University  No.19, Xinjiekouwai St, Haidian District, Beijing, 100875, P.R.China}
\author[0000-0003-2471-2363]{Haibo Yuan}
\author[0000-0003-1863-1268]{Ruoyi Zhang}
\affil{Department of Astronomy, Beijing Normal University No.19, Xinjiekouwai St, Haidian District, Beijing, 100875, P.R.China;  yuanhb@bnu.edu.cn}
\author[0000-0002-3651-0681]{Zexi Niu}
\affil{National Astronomical Observatories,
Chinese Academy of Sciences 
20A Datun Road, Chaoyang District,
Beijing, China}
\author[0000-0003-3250-2876]{Yang Huang}
\affil{South-Western Institute for Astronomy Research, Yunnan University, Kunming 650500, People’s Republic of China}
\author{Fuqing Duan}
\affil{College of Artificial Intelligence, Beijing Normal University No.19, Xinjiekouwai St, Haidian District, Beijing, 100875, P.R.China}
\author{Yi Fang}
\affil{Department of Astronomy, Beijing Normal University No.19, Xinjiekouwai St, Haidian District, Beijing, 100875, P.R.China;  yuanhb@bnu.edu.cn}

\begin{abstract}
In this letter, we have carried out an independent validation of the {\it Gaia} EDR3 photometry using about 10,000 Landolt standard stars from Clem \& Landolt (2013). Using a machine learning technique, the $UBVRI$ magnitudes are converted into the {\it Gaia} magnitudes and colors and then compared to those in the EDR3, with the effect of metallicity incorporated.
Our result confirms the significant improvements in the calibration process of the {\it Gaia} EDR3.
Yet modest trends up to 10 mmag with $G$ magnitude are found for all the magnitudes and colors
for the $10 < G < 19$ mag range, particularly for the bright and faint ends.
With the aid of synthetic magnitudes computed on the CALSPEC spectra with the {\it Gaia} EDR3 passbands,
absolute corrections are further obtained, paving the way for optimal usage of the {\it Gaia} EDR3 photometry 
in high accuracy investigations.
\end{abstract}


\section{INTRODUCTION}

%The {\it Gaia} Early Data Release 3 (Gaia Collaboration et al. 2020) has delivered not only the best astrometric
The Early Data Release 3 of the ESA's space mission {\it Gaia} ({\it Gaia} Collaboration et al. 2016, 2020)
has delivered not only the best astrometric
information but also the best photometric data for about 1.8 billion stars (Riello et al. 2020),
in terms of full sky coverage, uniform calibration at mmag level, and small photometric errors 
for a very wide range of magnitudes. However, due to the changes of instrument configurations, 
magnitude dependent systematic errors up to 10 mmag or higher have been detected in its 
DR2 (Riello et al. 2018; Casagrande \& VandenBerg 2018; Weiler 2018; 
Ma{\'\i}z Apell{\'a}niz \& Weiler 2018; Niu et al. 2021, re-submitted). 
Thanks to significant improvements in the calibration process, the magnitude term found in the 
{\it Gaia} DR2 photometry is greatly reduced in the EDR3. The 
overall trend is no larger than 1 mmag/mag except for very blue and bright sources (Riello et al. 2020). 
%($G_{\rm BP}-G_{\rm RP} < −0.1$), 
%where there exists a differential colour term between sources brighter and fainter than $G = 11$ 
%at a level of 1 per cent in magnitude. 
 
Due to the unprecedented photometric quality, it is challenging to identify possible problems of 
the {\it Gaia} photometry using external catalogs.
Synthetic magnitudes from well calibrated spectral libraries, such as the CALSPEC (Bohlin 2014), have been used to compare with the observed ones for the {\it Gaia} DR2 (Casagrande \& VandenBerg 2018; 
Weiler 2018; Ma{\'\i}z Apell{\'a}niz \& Weiler 2018). However, the number of available spectra is 
limited to a few hundreds, too few to identify any fine structures in the correction curves.
With about 0.5 million stars selected from the LAMOST DR5 (Luo et al. 2015), Niu et al. (2021, re-submitted) 
have applied the spectroscopy-based stellar color regression method (Yuan et al. 2015a) to calibrate the \textit{Gaia} 
DR2 $G - G_{\rm RP}$ and $G_{\rm BP}-G_{\rm RP}$ colors.
Systematic trends with $G$ magnitude are revealed for both colors in great detail at a precision of about 1 mmag.
However, contributions from each of the three \textit{Gaia} magnitudes can not be decoupled.
%effect of reddening 

In this letter, we aim to perform an independent test of the {\it Gaia} EDR3 photometry by comparing 
with the Landolt standard stars.
The high-quality CCD-based $UBVRI$ photometric data from Clem \& Landolt (2013; CL13 hereafter) 
is adopted for three reasons.
Firstly, it contains about 45,000 stars, about two orders of magnitude larger than the numbers of flux standards in spectral libraries. 
Secondly, it has a wide magnitude range ($10 < G < 20$) that matches well with the {\it Gaia} photometry. Last but not least, 
it has five filters ($UBVRI$), including the metallicity sensitive $U$ filter, making it possible to include 
the effect of metallicity when performing transformations between different photometric systems, 
which is essential but ignored in the official validation of the {\it Gaia} EDR3 (Riello et al. 2020; Fabricius et al. 2020).
Using a machine learning technique, the $UBVRI$ magnitudes are trained into the {\it Gaia} magnitudes and colors and then compared to those in the {\it Gaia} EDR3. Our result confirms the significant improvements in the calibration process of the {\it Gaia} EDR3. Yet modest trends with $G$ magnitude are found for all magnitudes and colors. 
By combining the synthetic magnitudes computed on the CALSPEC (Bohlin et al. 2014, 2020 April Update) spectra with the {\it Gaia} EDR3 passbands, 
we further obtain absolute corrections to the {\it Gaia} EDR3 photometry, paving the way for optimal usage of the {\it Gaia} photometry. 
 
The paper is organized as follows. In section 2, we introduce our data and method. The result is presented in Section 3 and discussed in Section 4. 
We summarize in Section 5. Note that in this paper, the \textit{Gaia} EDR3 $G$ magnitudes refer to the corrected ones (phot\_g\_mean\_mag\_corrected, Riello et al. 2020)

% ============================================================================================================
\section{DATA AND METHOD}\label{sect:data and method}

\begin{figure*}
    \centering
    \includegraphics[width=16cm]{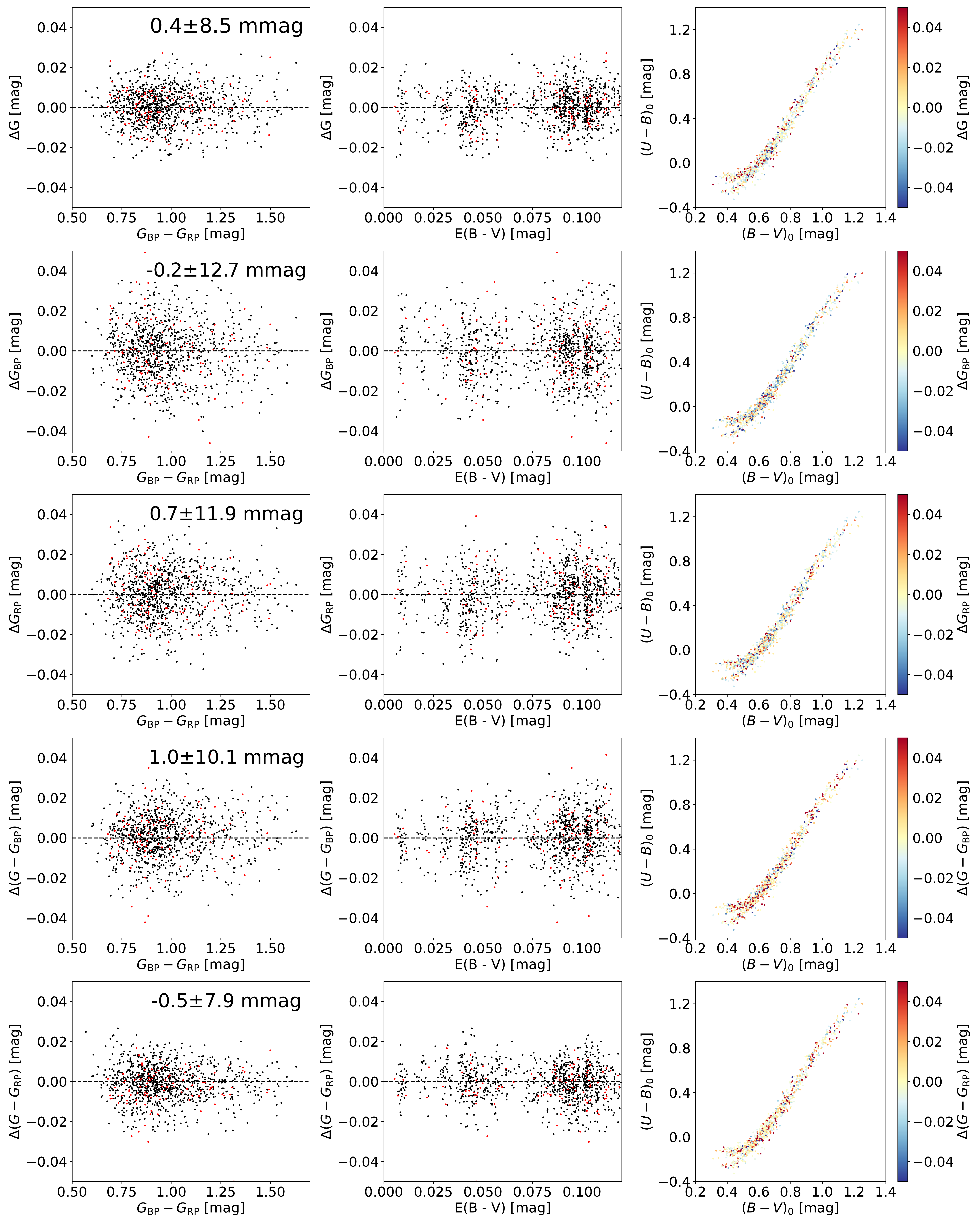}
    \caption{Residual distributions as functions of ${G}_{\rm BP} - {G}_{\rm RP}$ (left panel), 
reddening (middle panel),  and in the $(U-B)_{0} - (B-V)_{0}$ diagram (right panel) of the training (black dots) and test (red dots) samples for each of the five MLP networks. From top to bottom are the results for $G$, $G_{BP}$, $G_{RP}$, $G - G_{BP}$, and $G - G_{RP}$, respectively. The median values and standard deviations of the residuals are also marked in the left panels. 
For the reddening correction of the $U-B$ and $B-V$ colors, the SFD reddening map and reddening coefficients of 0.708 and 0.884 are adopted (Schlafly \& Finkbeiner 2011).}
    %    \emph{Top left}: comparison of LAMOST reddening estimates with those from the SFD98 map for selected individual stars.
    \label{f:fig1}
\end{figure*}

\subsection{Data}

In this work, we only use main sequence stars that have high-precision photometry from both the \textit{Gaia} EDR3 and the CL13. The following criteria are adopted to guarantee data quality: 

\begin{itemize}
   \item [1)]
    err($U$) $<$ 0.02 mag, err($B$) $<$ 0.015 mag, and err($V, R, I$) $<$ 0.01 mag to exclude stars of poor Landolt photometry; 
    \item [2)]
    \texttt{phot}$\_$\texttt{bp}$\_$\texttt{rp}$\_$\texttt{excess}$\_$\texttt{factor} $ < 1.26+0.04*(G_{\rm BP} - G_{\rm RP})^2$ to exclude stars of poor {\it Gaia} photometry (Evans et al. 2018). Note the above criterion is almost identical to requiring the 
    corrected \texttt{phot}$\_$\texttt{bp}$\_$\texttt{rp}$\_$\texttt{excess}$\_$\texttt{factor} (Riello et al. 2020) $<$ 0.08;
    \item [3)]
    $E(B–V)_{\rm SFD}$ $<$ 0.12 mag to exclude stars of high extinction, where $E(B–V)_{\rm SFD}$ is from the Schlegel et al. (1998, hereafter SFD) dust reddening map;
    \item [4)]
    $0.35<B-V<1.3$ mag to select F/G/K stars;
    \item [5)]
    $G_0>4*(G_{\rm BP-RP}-1.33*E(B-V)_{\rm SFD})$ to exclude giant stars, where $G_0$ is the intrinsic $G$ magnitude. Here we adopt an extinction coefficient of 2.5 for the $G$ band  and a reddening coefficient of 1.33 for the $G_{\rm BP} - G_{\rm RP}$ color (Chen et al. 2019). 
\end{itemize}  

Finally, 10,294 stars are selected, including 1,539 stars of $17 < G < 17.5$ mag as
the reference sample. The $17 < G < 17.5$  magnitude range is adopted in order to 1) have a large number of reference stars and 2) have a relatively wide range in metallicity. For the reference sample, we further exclude a few hundred stars whose magnitude errors are larger than 0.01 mag in the $U$ band or 0.005 mag in the $B/V/R/I$ band. The remaining 1,079 stars are divided into two groups: the training set (90 percent) and the test set (10 percent).  
For the whole sample, the median errors are 4.3, 3.3, 2.3, 1.9, and 2.3 mmag for $U,B,V,R,I$, and 2.8, 7.7, and 6.1 mmag for $G, G_{\rm BP}, G_{\rm RP}$, respectively. 
For the reference sample, the median errors are 3.6, 2.9, 1.9, 1.6, and 1.9 mmag for $U,B,V,R,I$, and 2.9, 9.2, and 7.1 mmag for $G, G_{\rm BP}, G_{\rm RP}$, respectively. 

\subsection{Method}

In this work, multi-layer perceptron neural networks (MLP) with architectures of 4-128-64-8-1 are designed to convert the Landolt $UBVRI$ photometry of CL13 into the {\it Gaia} EDR3 magnitudes and colors. Each node of the hidden layers and the output layer is connected to all nodes of its previous layer, with a nonlinear function:
\begin{equation} 
    f(X)=g(WX+b)
\end{equation}
where $W$ and $b$ respectively represent the weight matric and bias vector, $g(\bullet)$ represents the LeakyReLU activation function with negative slope coefficient $\alpha =0.1$ in the hidden layer.
Five MLPs are trained, three for the {\it Gaia} EDR3 magnitudes and two for the {\it Gaia} EDR3 colors.
%Each MLP maps the inputs to a predicted variable 
The four input colors are the same for the five MLPs, i.e., $U-B, B-V, V-R, R-I$.
The outputs are $G-A$, ${G}_{\rm BP} - A$,  ${G}_{\rm RP} - A$, $G-{G}_{\rm BP}$, 
and ${G}-{G}_{\rm RP}$ respectively, where $A$ is an artificial magnitude defined as the mean of the $B, V, R$ magnitudes. Note that in this work, observed magnitudes/colors are preferred to 
dereddened ones, to avoid possible systematic errors caused by reddening correction. 
Systematic errors may come from at least two aspects: spatially dependent systematic errors with 
the SFD reddening map (Sun et al., to be submitted; Niu et al. 2021) and overestimates of reddening 
for bright local stars.  

The training process is carried out with Keras 2.2.4 and Tensorflow 1.12. The loss function, Mean Squared Error (MSE) with a $L_2$ regularized term, is optimized using adaptive moment estimation (ADAM, Kingma \& Ba 2014) with a mini-batch size of 900 samples. The trade-off coefficient between the MSE and regularized term is 0.000001 to avoid overfitting. Other hyper-parameters set manually in our work are training iterations $epoch=100,000$ and learning rate $\eta =0.00001$.  

During the training process, a $3 \sigma$  clipping is performed to exclude outliers. Then, the training process runs again with the same hyper-parameters. % as the previous model.
After the networks are well trained, we apply the models to the whole dataset. The predicted magnitudes and colors are then obtained and compared to those in the EDR3. The median differences (predicted $-$ observed) as a function of $G$ magnitude are regarded as the calibration curves.
%Corrections can be described as:
%\begin{gather}
%  {Mag}_{corrected}=Mag+\Delta M \\
%  {Color}_{corrected}=Color+\Delta C
%\end{gather}

\section{Result} % towards the M31 region 
Figure 1 shows the results of the training and test samples for different MLP neural networks.
It can be seen that the residuals show no dependence on the ${G}_{\rm BP} - {G}_{\rm RP}$ color or the SFD reddening. The residuals also show no systematic patterns in the $(U-B)_{0} - (B-V)_{0}$ diagram. Because dwarf stars of different metallicities are well separated in the $(U-B)_{0} - (B-V)_{0}$ diagram (e.g., Sandage \& Smith 1963), the results suggest that the effect of metallicity has been fully taken into account in our neural networks. Note that the standard deviations of the residuals are 8.5, 12.7, 11.9, 10.1, and 7.9 mmag for the $G$, $G_{\rm BP}$, $G_{\rm RP}$, $G - G_{\rm BP}$, and $G - G_{\rm RP}$, respectively. The small standard deviations suggest
that the {\it Gaia} photometry can be well recovered from the Landolt photometry, 
to a precision of about 1 percent with photometric errors included.

Figure 2 shows the residual distributions in the ($G_{\rm BP} - G_{\rm RP}$) -- $G$ diagram 
for the whole dataset. No obvious dependence on color within the two dashed lines is seen for all
the panels, consistent with the result of Riello et al. (2020). For a few stars of $G_{\rm BP} - G_{\rm RP} < 0.6$ or $G_{\rm BP} - G_{\rm RP} > 1.5$, there exist some discrepancies, probably caused by the boundary effect in the training process. Those stars are excluded in the following analysis. 

\begin{figure*}
    \centering
    \includegraphics[width=\linewidth]{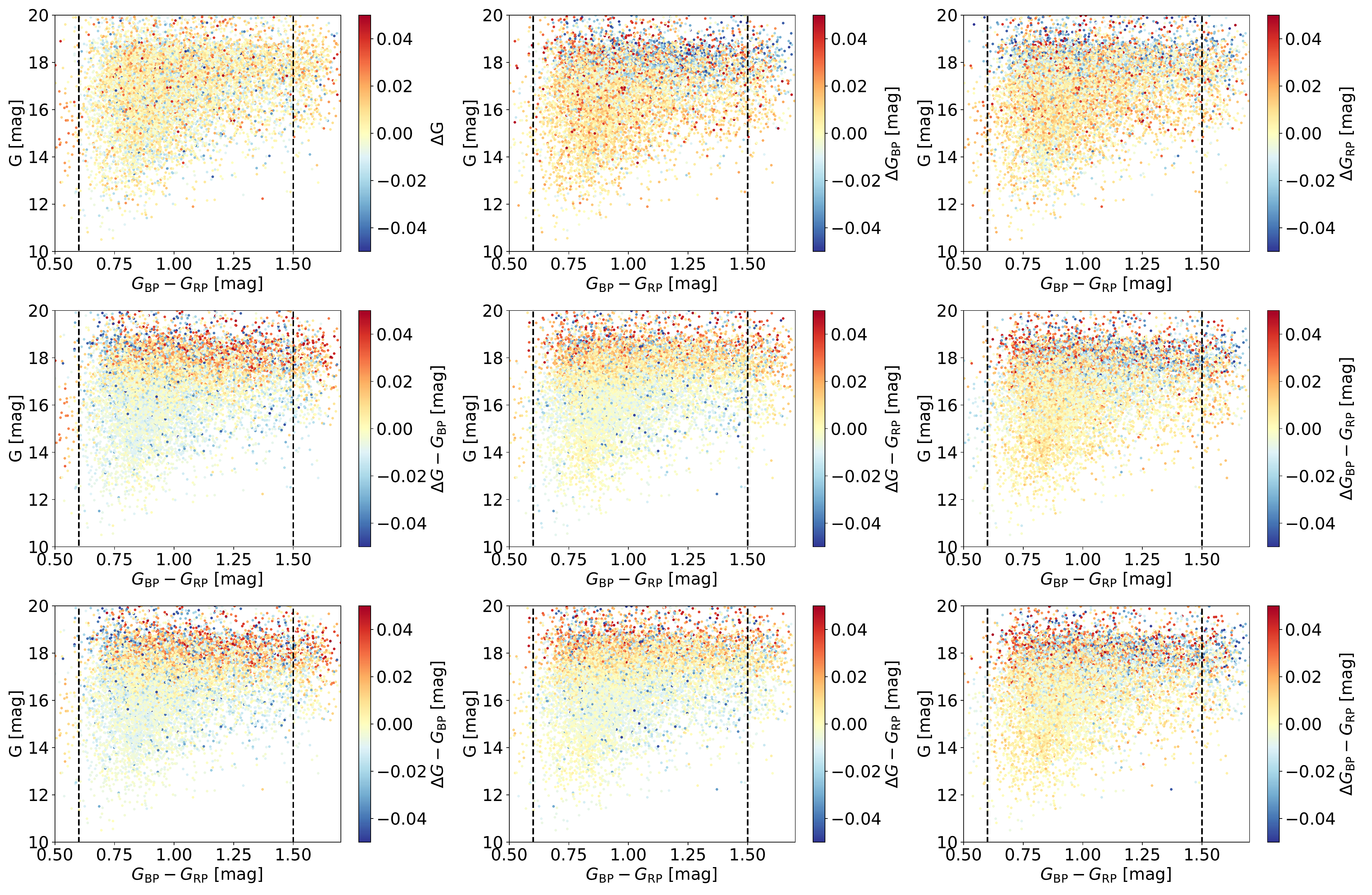}
    \caption{Residual distributions in the ($G_{\rm BP} - G_{\rm RP}$) -- $G$ diagram for the whole dataset. 
The top panels show the results of the {\it Gaia} magnitudes.
The middle panels show the results of the {\it Gaia} colors calculated from the top panels.
The two left panels in the bottom show the results of the {\it Gaia} colors directly trained from the neural networks, while the right one the result calculated from the two left panels. The two vertical dashed lines indicate $G_{BP} - G_{RP}$ color of 0.6 and 1.5 mag, respectively.\label{f:fig2}}
\end{figure*}

\begin{figure*}
    \centering
    \includegraphics[width=\linewidth]{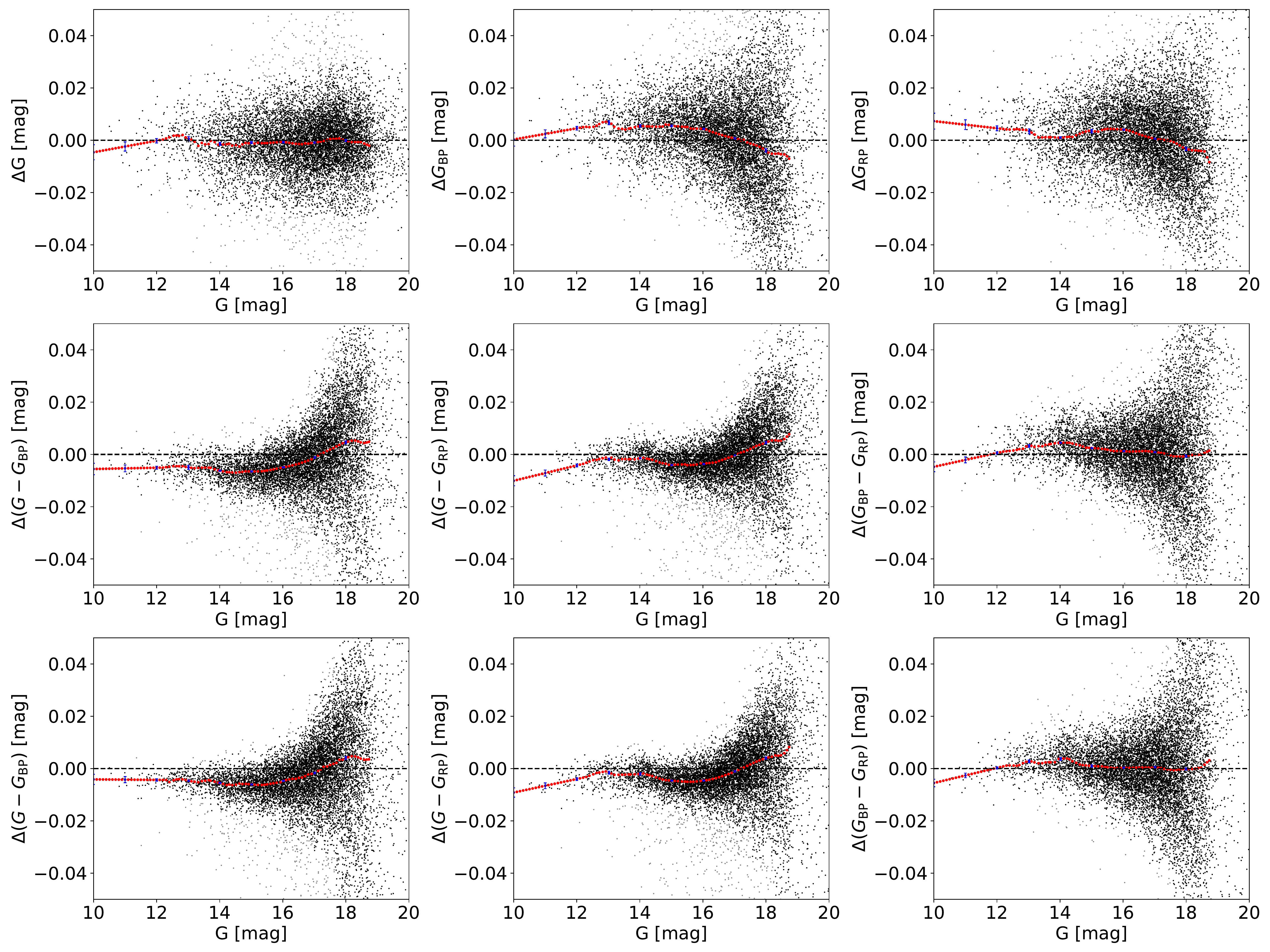}
    \caption{Residual distributions as a function of $G$ magnitude for the whole dataset after excluding stars of  $G_{\rm BP} - G_{\rm RP} < 0.6$ and $G_{\rm BP} - G_{\rm RP} > 1.5$. 
The panels are arranged in the same way to Figure 2. 
For each panel, stars are divided into different bins of width of 0.4 mag at a step size 
of 0.1 mag when $12.5 < G < 18.7$ mag.  The median value for each bin is estimated,  
with a $3-\sigma$ clipping  performed and the grey dots dropped. At the bright end, 
a linear fitting is performed for stars of $G < 13.0$ mag, the results are adopted when $G < 12.5$ mag.
Finally, a locally estimated scatterplot smoothing (LOWESS) is applied to smooth the calibration curves, 
with $frac = 0.07$. The final results are indicated by red dotted lines. 
The blue error bars are estimated with 500 subsamples using the Bootstrap method.
\label{f:fig3}}
\end{figure*}

The calibration curves as a function of $G$ magnitude for the  {\it Gaia} magnitudes and colors 
are plotted in Figure 3. The errors are also estimated using the Bootstrap method with 500 subsamples. 
The errors at G $\sim$ 12.0, 13.0, 14.0, 16.0, and 18.0 mag are respectively
(0.9, 0.8, 0.6, 0.4, 0.4), (0.7, 0.7, 0.5, 0.3, 0.7), and (0.9, 0.9, 0.3, 0.3, 0.6) mmag in the top panels from left to right,
(0.6, 0.8, 0.2, 0.2, 0.6), (0.6, 0.4, 0.2, 0.2, 0.5), and (0.5, 0.5, 0.2, 0.2, 0.5) mmag in the middle panels from left to right,
(0.5, 0.3, 0.2, 0.2, 0.6), (0.5, 0.6, 0.4, 0.2, 0.4), and (0.4, 0.5, 0.5, 0.3, 0.4) mmag in the bottom panels  from left to right. 
Our result confirms the significant improvements in the calibration process of the {\it Gaia} EDR3. The strong trend in $G$ as a function of $G$ in DR2 is greatly reduced. 
Yet modest trends with $G$ magnitude are found for all magnitudes and colors. 
A tiny discontinuity of 2--3 mmag at $G \sim 13$ mag is clearly detected for all the magnitude curves, 
probably related to a change in the instrument configuration. % corresponding to the transition between window class 0 and 1 (Gaia Collaboration et al. 2016).
The downturn at faint magnitudes visible in the $G_{\rm BP}$ and $G_{\rm RP}$ passbands is possibly caused by some over-estimation of the background in the BP and RP spectra.
Therefore, the trend in $G_{\rm BP} - G_{\rm RP}$ as a function of $G$ is much weaker compared to that 
in $G - G_{\rm BP}$ and $G - G_{\rm RP}$.

In the above analysis, we have assumed that the corrections are zero for stars
of $17 < G < 17.5$ mag, which is not true. 
To set an absolute correction zero point, synthetic magnitudes of $G$, $G_{\rm BP}$, and $G_{\rm RP}$ are computed on the CALSPEC (Bohlin et al. 2014, 2020 April Update) spectra with 
the {\it Gaia} EDR3 passbands, with the same approach of Riello et al. (2020).  The results 
are used to obtain the absolute corrections.%, with a minimum $\chi^2$ method. 
Only 20 stars of $G_{BP} - G_{RP} > -0.3$, $G > 10$, and \texttt{phot}$\_$\texttt{bp}$\_$\texttt{rp}$\_$\texttt{excess}$\_$\texttt{factor} $ < 1.26+0.04*(G_{\rm BP} - G_{\rm RP})^2$ are used. 
%The shifts are typically below 0.01 mag, with formal random errors of about 2 mmag.
The mean magnitude offsets are $-$4.2, $-$9.5, and 3.0 mmag for $G$, $G_{\rm BP}$, and $G_{\rm RP}$, respectively.
The shifts of the calibration curves are 
$-$2.1, $-$15.1, $-$1.0, 9.9, and $-$3.1 mmag for $G$, $G_{\rm BP}$, $G_{\rm RP}$, $G - G_{\rm BP}$, 
and $G - G_{\rm RP}$, respectively.% The absolute values are much larger for $G_{\rm BP}$ and $G - G_{\rm BP}$, 
%compared to those of other magnitudes and colors.}
The final calibration curves are plotted in Figure 4 and listed in Table 1.
Note that only results directly obtained from the five MLPs (red lines in the top panels of Figure 4 and 
blue lines in the two bottom panels on the left) are given in Table 1.
Calibration curves yielded by different MLPs (blue and red dotted lines in Figure 4) agree with 
each other very well, with a typical difference of 0.5 mmag, within the training errors. 

%Note that the EDR3 photometry is calibrated in the absolute flux level using a set of 
%spectro-photometric standard stars (SPSS;  Pancino et al. 2012), whose fluxes are tied to the 
%2013 version of CALSPEC (Riello et al. 2020). Adjusting the absolute flux scale to the current CALSPEC 
%library will therefore result in an inconsistency with the uncorrected photometry. 

%The calibration with respect to CALSPEC is somewhat problematic: The EDR3 photometry is calibrated in the absolute flux level using a set of calibration spectra (the "SPSS"), which itself is (rather loosely) related to the CALSPEC pillars in their Jan 2010 version. Adjusting the absolute flux scale to the current CALSPEC library will therefore result in an inconsistency with the uncorrected photometry. The authors should (briefly) discuss such inconsistencies, and specify the mean offset derived from the CALSPEC synthetic photometry, and the value of the resulting discontinuity at G=10mag.

\begin{figure*}
    \centering
    \includegraphics[width=\linewidth]{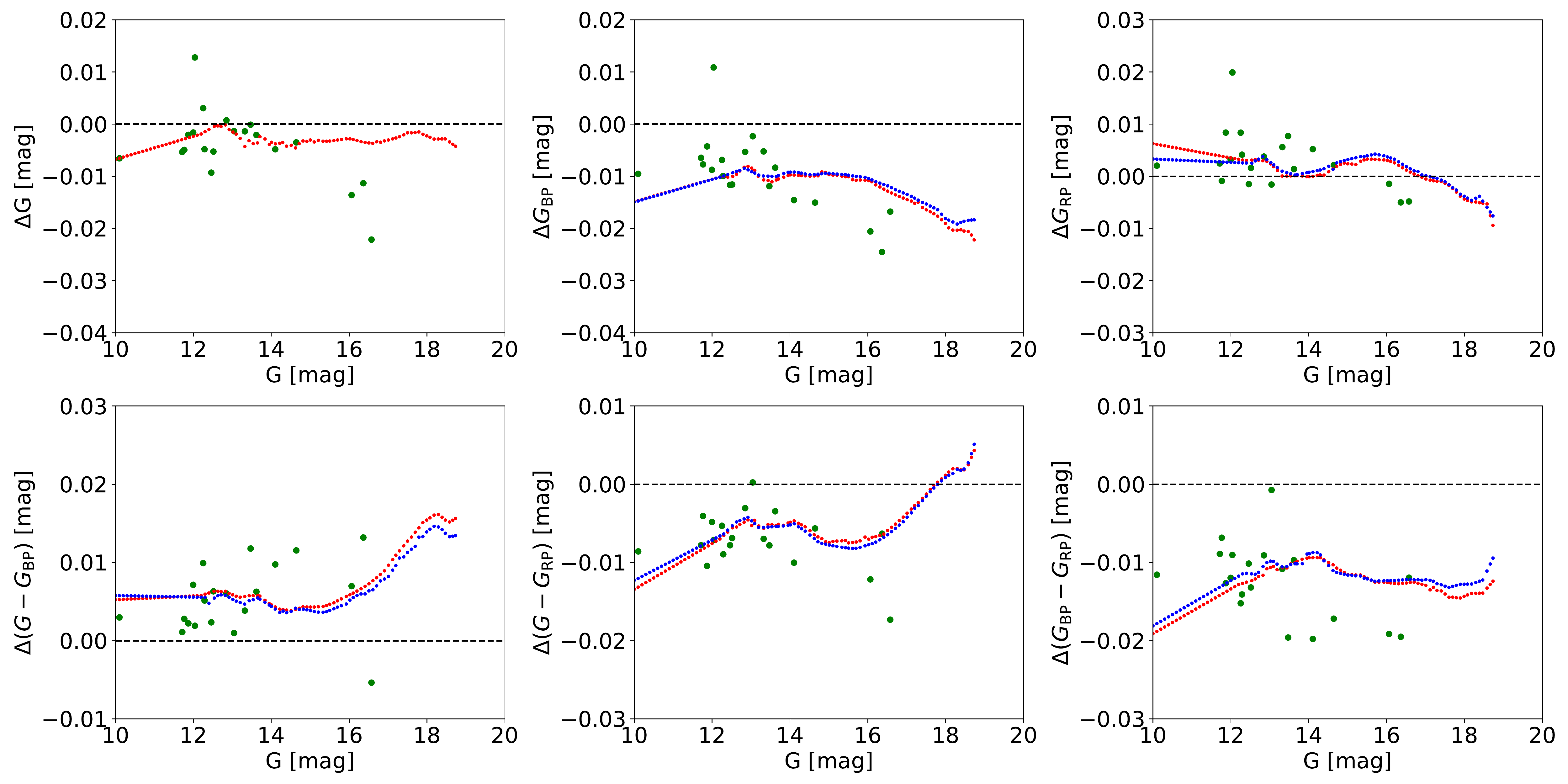}
    \caption{Magnitude (top panels) and color (bottom panels) calibration curves. The green dots are results from synthetic magnitudes/colors of the CALSPEC spectra. The red dotted lines 
    are results from the trained magnitudes. The blue dotted lines in the two top panels on the right are results combining the trained $G$ magnitude and trained $G - G_{\rm BP}$ and  $G - G_{\rm RP}$ colors. 
    The blue dotted lines in the bottom panels are results from the trained colors. All the dotted lines are shifted by the median of the differences between the dotted lines and the green dots to match the green dots.\label{f:fig4}}
\end{figure*}

\begin{table}[htbp]
    \centering
    \caption{Magnitude and color calibration curves in units of mmag.}
    \label{table:tab1}
    \begin{tabular}{p{0.85cm}<{\centering}p{0.9cm}<{\centering}p{1.05cm}<{\centering}p{1.05cm}<{\centering}p{1.25cm}<{\centering}p{1.25cm}<{\centering}}
    \hline
     $G$ &  $\Delta G$ &  $\Delta G_{\rm BP}$ &  $\Delta G_{\rm RP}$ & $\Delta (G-G_{\rm BP})$ & $\Delta (G-G_{\rm RP})$\\
    \hline 
    10.0  &    $-$6.7  &    $-$14.9  &    6.3  &    5.8  &    $-$12.3  \\
    10.4  &    $-$5.9  &    $-$14.0  &    5.7  &    5.7  &    $-$11.3  \\
    10.8  &    $-$5.0  &    $-$13.1  &    5.2  &    5.7  &    $-$10.2  \\
    11.2  &    $-$4.1  &    $-$12.3  &    4.6  &    5.7  &    $-$9.2  \\
    11.6  &    $-$3.2  &    $-$11.4  &    4.1  &    5.6  &    $-$8.2  \\
    12.0  &    $-$2.3  &    $-$10.6  &    3.5  &    5.6  &    $-$7.1  \\
    12.4  &    $-$1.0  &    $-$10.1  &    3.1  &    4.8  &    $-$5.9  \\
    12.5  &    $-$0.6  &    $-$10.0  &    3.1  &    5.3  &    $-$5.4  \\
    12.6  &    $-$0.4  &    $-$9.7  &    3.2  &    5.7  &    $-$5.0  \\
    12.7  &    $-$0.4  &    $-$9.1  &    3.1  &    5.8  &    $-$4.7  \\
    12.8  &    $-$0.2  &    $-$8.4  &    3.0  &    5.8  &    $-$4.5  \\
    12.9  &    $-$0.9  &    $-$8.1  &    2.9  &    5.6  &    $-$4.3  \\
    13.0  &    $-$1.4  &    $-$8.4  &    2.5  &    5.3  &    $-$4.6  \\
    13.1  &    $-$1.9  &    $-$8.9  &    1.9  &    5.1  &    $-$5.1  \\
    13.2  &    $-$2.7  &    $-$10.0  &    1.1  &    4.9  &    $-$5.5  \\
    13.3  &    $-$4.0  &    $-$10.6  &    0.2  &    4.7  &    $-$5.5  \\
    13.4  &    $-$3.5  &    $-$10.8  &    0.1  &    5.0  &    $-$5.5  \\
    13.5  &    $-$3.5  &    $-$11.0  &    0.1  &    5.2  &    $-$5.5  \\
    13.6  &    $-$3.6  &    $-$10.8  &    0.1  &    5.4  &    $-$5.4  \\
    13.7  &    $-$2.6  &    $-$10.5  &    0.1  &    5.3  &    $-$5.4  \\
    13.8  &    $-$2.7  &    $-$10.3  &    0.1  &    5.0  &    $-$5.3  \\
    13.9  &    $-$3.4  &    $-$10.0  &    0.0  &    4.7  &    $-$5.3  \\
    14.0  &    $-$3.5  &    $-$9.7  &    $-$0.1  &    4.4  &    $-$5.2  \\
    14.1  &    $-$3.7  &    $-$9.7  &    0.0  &    4.1  &    $-$5.1  \\
    14.2  &    $-$3.7  &    $-$9.8  &    0.2  &    3.7  &    $-$5.4  \\
    14.3  &    $-$3.5  &    $-$9.9  &    0.3  &    3.8  &    $-$5.7  \\
    14.4  &    $-$4.2  &    $-$9.9  &    0.3  &    3.6  &    $-$6.0  \\
    14.5  &    $-$4.1  &    $-$10.0  &    0.8  &    3.7  &    $-$6.4  \\
    14.6  &    $-$4.4  &    $-$9.9  &    1.3  &    4.1  &    $-$6.9  \\
    14.7  &    $-$3.9  &    $-$9.9  &    1.8  &    4.0  &    $-$7.3  \\
    14.8  &    $-$3.3  &    $-$9.3  &    2.2  &    4.0  &    $-$7.5  \\
    14.9  &    $-$3.3  &    $-$9.2  &    2.5  &    4.0  &    $-$7.6  \\
    15.0  &    $-$3.0  &    $-$9.7  &    2.4  &    3.9  &    $-$7.8  \\
    15.1  &    $-$3.4  &    $-$9.8  &    2.4  &    3.7  &    $-$7.9  \\
    15.2  &    $-$3.1  &    $-$9.8  &    2.3  &    3.7  &    $-$7.9  \\
    15.3  &    $-$3.2  &    $-$9.9  &    2.7  &    3.6  &    $-$8.0  \\
    15.4  &    $-$3.3  &    $-$10.0  &    3.1  &    3.7  &    $-$8.1  \\
    \hline
    \end{tabular}
    \end{table}
\begin{table}[htbp]
    \centering
    \begin{tabular}{p{0.85cm}<{\centering}p{0.9cm}<{\centering}p{1.05cm}<{\centering}p{1.05cm}<{\centering}p{1.25cm}<{\centering}p{1.25cm}<{\centering}}
    \hline
    $G$ &  $\Delta G$ &  $\Delta G_{\rm BP}$ &  $\Delta G_{\rm RP}$ & $\Delta (G-G_{\rm BP})$ & $\Delta (G-G_{\rm RP})$\\
    \hline 	
    15.5  &    $-$3.2  &    $-$10.1  &    3.3  &    3.9  &    $-$8.2  \\
    15.6  &    $-$3.1  &    $-$10.6  &    3.3  &    4.1  &    $-$8.2  \\
    15.7  &    $-$3.0  &    $-$10.8  &    3.3  &    4.3  &    $-$8.2  \\
    15.8  &    $-$2.9  &    $-$10.7  &    3.2  &    4.5  &    $-$8.1  \\
    15.9  &    $-$2.8  &    $-$10.7  &    3.2  &    4.6  &    $-$7.9  \\
    16.0  &    $-$2.8  &    $-$10.8  &    3.1  &    5.1  &    $-$7.8  \\
    16.1  &    $-$2.9  &    $-$11.0  &    2.9  &    5.5  &    $-$7.6  \\
    16.2  &    $-$3.1  &    $-$11.6  &    2.6  &    5.8  &    $-$7.4  \\
    16.3  &    $-$3.3  &    $-$12.0  &    2.2  &    6.0  &    $-$7.1  \\
    16.4  &    $-$3.5  &    $-$12.4  &    1.7  &    6.0  &    $-$6.8  \\
    16.5  &    $-$3.6  &    $-$12.8  &    1.2  &    6.3  &    $-$6.5  \\
    16.6  &    $-$3.7  &    $-$13.2  &    0.8  &    6.5  &    $-$6.1  \\
    16.7  &    $-$3.4  &    $-$13.5  &    0.5  &    7.0  &    $-$5.7  \\
    16.8  &    $-$3.5  &    $-$13.8  &    0.1  &    7.6  &    $-$5.3  \\
    16.9  &    $-$3.2  &    $-$14.1  &    $-$0.2  &    7.9  &    $-$4.8  \\
    17.0  &    $-$3.0  &    $-$14.5  &    $-$0.5  &    8.2  &    $-$4.3  \\
    17.1  &    $-$2.8  &    $-$14.7  &    $-$0.7  &    8.9  &    $-$3.7  \\
    17.2  &    $-$2.7  &    $-$15.2  &    $-$0.8  &    9.6  &    $-$3.1  \\
    17.3  &    $-$2.4  &    $-$15.0  &    $-$0.9  &    10.6  &    $-$2.7  \\
    17.4  &    $-$2.1  &    $-$16.0  &    $-$1.0  &    10.7  &    $-$2.1  \\
    17.5  &    $-$1.7  &    $-$16.4  &    $-$1.3  &    11.3  &    $-$1.5  \\
    17.6  &    $-$1.6  &    $-$16.8  &    $-$1.8  &    11.7  &    $-$1.0  \\
    17.7  &    $-$1.6  &    $-$17.2  &    $-$2.3  &    12.1  &    $-$0.4  \\
    17.8  &    $-$1.5  &    $-$17.7  &    $-$3.1  &    13.2  &    0.0  \\
    17.9  &    $-$1.9  &    $-$18.4  &    $-$3.8  &    13.3  &    0.5  \\
    18.0  &    $-$2.3  &    $-$19.1  &    $-$4.4  &    13.9  &    0.9  \\
    18.1  &    $-$2.6  &    $-$20.0  &    $-$4.7  &    14.3  &    1.2  \\
    18.2  &    $-$2.9  &    $-$20.3  &    $-$4.9  &    14.6  &    1.4  \\
    18.3  &    $-$2.8  &    $-$20.3  &    $-$4.9  &    14.5  &    1.9  \\
    18.4  &    $-$2.8  &    $-$20.3  &    $-$5.1  &    14.1  &    1.8  \\
    18.5  &    $-$3.0  &    $-$20.5  &    $-$5.2  &    13.6  &    2.2  \\
    18.6  &    $-$3.5  &    $-$20.8  &    $-$5.9  &    13.3  &    3.0  \\
    18.7  &    $-$4.0  &    $-$21.8  &    $-$8.6  &    13.4  &    4.6  \\
    \hline
    \end{tabular}
    \end{table}

\section{DISCUSSION} 

Fabricius et al. (2020) compare {\it Gaia} EDR3 photometry to a number of external catalogs in their Figure 34, including the one we use in the current work. 
They select all stars of $|b| > 30\degree$ and $A_V < 0.05$ mag for low latitude stars. 
Simple color-color relations are then used, $X = V + f(V - I)$ where $X$ denotes {\it Gaia} magnitude,  to transform the Landolt $V,I$ magnitudes into {\it Gaia} magnitudes. 

Stellar colors depend mainly on the stellar effective temperature, but also to a fair degree on metallicity, particularly the blue colors.
Yuan et al. (2015) propose to use the 
metallicity-dependent stellar locus to better describe the transformation relations between different colors. Taking the SDSS colors for example, at a given $g-i$ color, they find that typically 1 dex decrease in metallicity causes 0.20 and 0.02 mag decrease in colors $u - g$ and $g-r$ and 0.02 and 0.02 mag increase in colors $r-i$ and $i-z$, respectively. The variations are larger for more metal-rich stars, and for F/G/K stars. The relations are also different between dwarf stars and giant stars. Therefore, to make optimal transformations between different colors, the metallicity effect shall be taken into account. 
Huang et al. (2020) have applied a revised stellar color regression method to re-calibrate the SkyMapper Southern Survey DR2 (Onken et al. 2019), achieving a uniform calibration with precision better than 1 percent by considering the 
effect of metallicity on stellar colors. 
López-Sanjuan et al. (2021) have discussed the impact of metallicity on photometric calibration of the 
JPLUS survey (Cenarro et al. 2019) with the stellar locus technique and found significant improvements for blue passbands.

A large vertical metallicity gradient of the Galaxy at the solar neighborhood is widely reported, e.g., 0.15 dex kpc$^{-1}$ from Huang et al. (2015). Considering the strong correlation
between stellar magnitudes and distances for dwarf stars, it implies a strong correlation between stellar magnitudes and metallicities, especially for high Galactic latitude regions.
Such a correlation could cause magnitude dependent systematic errors when adopting simple color-color relations and ignoring the effect of metallicity, at the level of from several mmag to tens of mmag for the {\it Gaia} passbands, depending on which colors are used and properties of the sample stars.  
In our work, by making use of the full color information of the Landolt photometry, particularly the $U-B$ color,  we can naturally incorporate the effect of metallicity and obtain robust results.

\begin{figure}
    \centering
    \includegraphics[width=\linewidth]{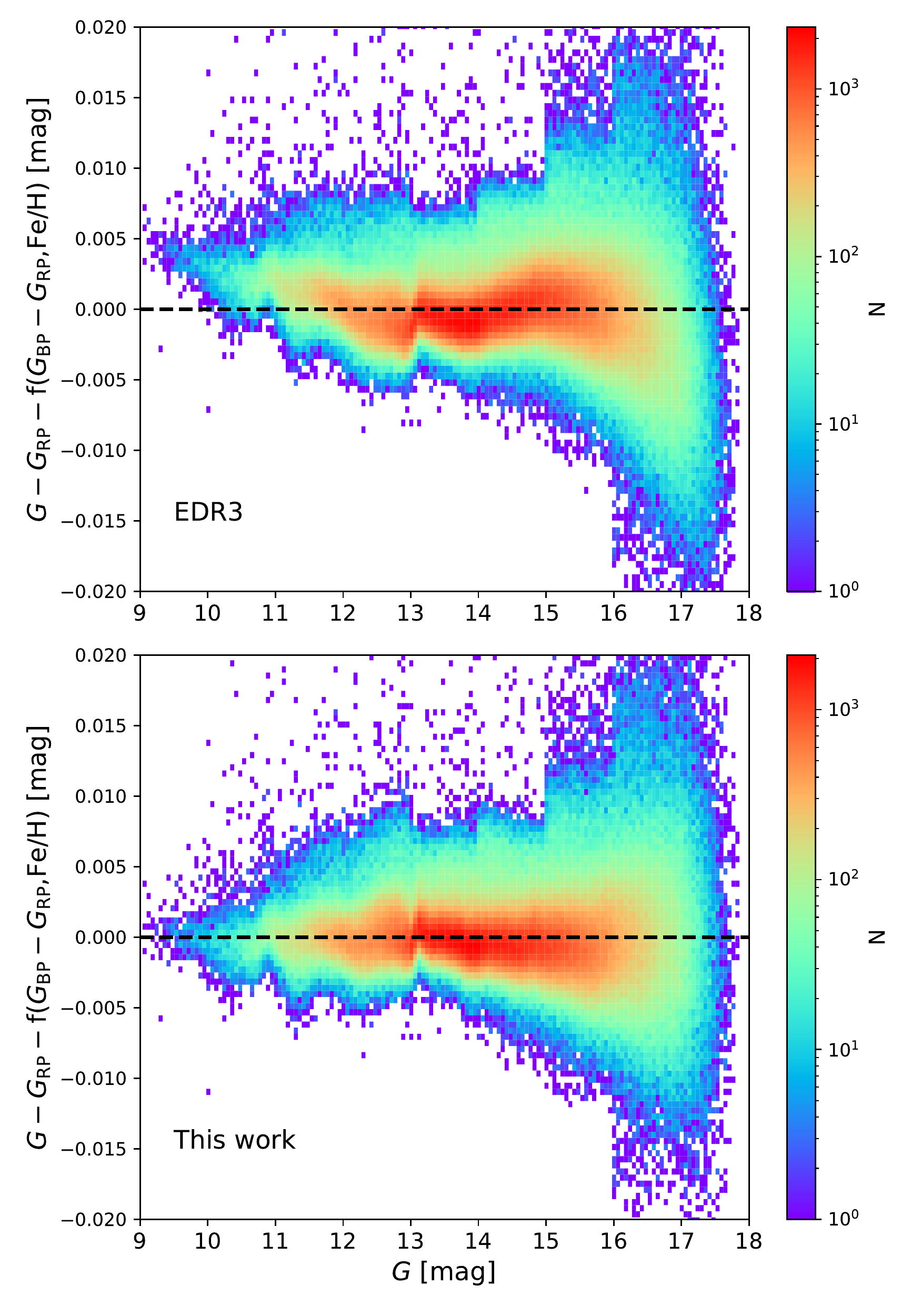}
    \caption{2D histogram of the $G - G_{\rm BP}$ residual after subtracting the metallicity-dependent color locus. Top: published EDR3 data. Bottom: applying magnitude corrections from this work. Note that the colors here refer to dereddened colors. \label{f:fig5}}
\end{figure}

To validate our result, following the procedure in Niu et al. (2021, re-submitted), we select a high-quality common sample containing 0.7 million stars in LAMOST DR7 and \textit{Gaia} EDR3 and plot the residuals from the $G - G_{\rm RP} = f(G_{\rm BP} - G_{\rm RP}, \rm{[Fe/H]})$ relation as a function of $G$ in Figure \ref{f:fig5}. The top panel plots the result of the published EDR3 data, which is quite similar to Figure 32 in Fabricius et al. (2020) and shows small but well-detected magnitude dependent deviations from zero. The bottom panel plots the result after the magnitude corrections in this work. Compared to the published EDR3 data, the deviations are significantly reduced to zero along with the $G$ magnitude, especially at the bright and faint ends. The result suggests that our corrections are valid. Note that at $G \sim 13$ mag where a discontinuity happens, our corrections are not as good as for fainter magnitudes. This is because we do not have enough stars to sample the correction curves at a very high resolution in $G$ magnitude.

Note that due to the limited magnitude and color ranges ($10 < G < 19$, $0.6 < G_{\rm BP} - G_{\rm RP} < 1.5$) 
of our final sample, our correction curves may not be valid for stars outside the above ranges, particularly for the bright and
blue ($G < 13$ and $G_{\rm BP}-G_{\rm RP} < -0.1$) ones (Riello et al. 2020), and thus should be used with caution.  
Note also that the EDR3 photometry is calibrated in the absolute flux level using a set of 
spectro-photometric standard stars (SPSS;  Pancino et al. 2012), whose fluxes are tied to the 
2013 version of CALSPEC (Riello et al. 2020). Adjusting the ab
solute flux scale to the current CALSPEC 
library will therefore result in an inconsistency with the uncorrected photometry. 
To avoid discontinuities, the calibration terms at $G=10$ are suggested for stars brighter than $G = 10$. 
For stars fainter than $G = 18.7$, the calibration terms at $G=18.7$ are suggested.

In the current MLPs, reddening values of individual stars are not included. We have performed tests by including reddening values from the SFD map into the networks, and the predicted magnitudes and colors are hardly changed, suggesting that our results are insensitive to the reddening. It is not surprising because systematic errors in the reddening correction are largely canceled out.

\section{SUMMARY}

In this work, we have carried out an independent validation of the {\it Gaia} EDR3 photometry using about 10,000 well selected Landolt standard stars from Clem \& Landolt (2013). 
Using five MLPs with architectures of 4-128-64-8-1, the $UBVRI$ magnitudes are trained into the {\it Gaia} magnitudes and colors and then compared to those in the {\it Gaia} EDR3,  with the effect of metallicity fully taken into account.

Our result confirms the significant improvements in the calibration process of the {\it Gaia} EDR3. The strong trend in $G$ as a function of $G$ in DR2 is greatly reduced. 
Yet modest trends with $G$ magnitude are found for all magnitudes and colors for the $10 < G < 19$ mag range, particularly at the bright and faint ends.
A tiny discontinuity of 2--3 mmag at $G \sim 13$ mag is clearly detected for all the magnitude curves, 
probably related to a change in the instrument configuration. 
The downturn at faint magnitudes visible in the $G_{\rm BP}$ and $G_{\rm RP}$ passbands is possibly caused by some over-estimation of the background in the BP and RP spectra.
The trend in $G_{\rm BP} - G_{\rm RP}$ as a function of $G$ is much weaker compared to that in $G - G_{\rm BP}$ and $G - G_{\rm RP}$.
With synthetic magnitudes computed on the CALSPEC spectra with the {\it Gaia} EDR3 passbands,
absolute calibration curves are further provided (Figure 4 and Table 1), paving the way for optimal usage of the {\it Gaia} EDR3 photometry in high accuracy investigations.
In the future {\it Gaia} data releases, the effect of metallicity should be included when comparing with external catalogs.

Our result demonstrates that mapping from one set of observables directly to another set of observables 
with machine learning provides a promising way in the calibration and analyses of large scale surveys.
%future Gaia, include feh

\begin{acknowledgements}
We acknowledge the anonymous referee for his/her valuable comments that improve the quality of this paper.
%We acknowledge very helpful discussions with Xiaowei Liu, Biwei Jiang, Jiang Gao, Bingqiu Chen, Maosheng Xiang, and Yang Huang.
This work is supported by the National Natural Science Foundation of China through the projects NSFC 11603002, the National Key Basic R\&D Program of China via 2019YFA0405503, and Beijing Normal University grant No.310232102.
This work has made use of data from the European Space Agency (ESA) mission {\it Gaia} (https://www.cosmos.esa.int/gaia), processed by the {\it Gaia} Data Processing and Analysis 
Consortium (DPAC, https://www.cosmos.esa.int/
web/gaia/dpac/ consortium). Funding for the DPAC has been provided by national institutions, in particular the institutions participating in the {\it Gaia} Multilateral Agreement.
Guoshoujing Telescope (the Large Sky Area Multi-Object Fiber Spectroscopic Telescope LAMOST) is a National Major Scientific Project built by the Chinese Academy of Sciences. Funding for the project has been provided by the National Development and Reform Commission. LAMOST is operated and managed by the National Astronomical Observatories, Chinese Academy of Sciences.
This research has made use of the SIMBAD database,
operated at CDS, Strasbourg, France.
\end{acknowledgements}

% ============================================================================================================
%\bibliography{ref}

\begin{thebibliography}{4}                                                                                                                                                 
\providecommand\natexlab[1]{#1}
\providecommand\JournalTitle[1]{#1}

\bibitem[Bohlin et al.(2014)]{2014PASP..126..711B} Bohlin, R.~C., Gordon, K.~D., \& Tremblay, P.-E.\ 2014, \pasp, 126, 711. doi:10.1086/677655

\bibitem[Casagrande \& VandenBerg(2018)]{2018MNRAS.479L.102C} Casagrande, L. \& VandenBerg, D.~A.\ 2018, \mnras, 479, L102. doi:10.1093/mnrasl/sly104

\bibitem[Cenarro et al.(2019)]{2019A&A...622A.176C} Cenarro, A.~J., Moles, M., Crist{\'o}bal-Hornillos, D., et al.\ 2019, \aap, 622, A176. doi:10.1051/0004-6361/201833036

\bibitem[Chen et al.(2019)]{2019MNRAS.483.4277C} Chen, B.-Q., Huang, Y., Yuan, H.-B., et al.\ 2019, \mnras, 483, 4277. doi:10.1093/mnras/sty3341

\bibitem[Clem \& Landolt(2013)]{2013AJ....146...88C} Clem, J.~L. \& Landolt, A.~U.\ 2013, \aj, 146, 88. doi:10.1088/0004-6256/146/4/88

\bibitem[Evans et al.(2018)]{2018A&A...616A...4E} Evans, D.~W., Riello, M., De Angeli, F., et al.\ 2018, \aap, 616, A4. doi:10.1051/0004-6361/201832756

\bibitem[Fabricius et al.(2020)]{2020arXiv201206242F} Fabricius, C., Luri, X., Arenou, F., et al.\ 2020, arXiv:2012.06242

\bibitem[Gaia Collaboration et al.(2016)]{2016A&A...595A...1G} Gaia Collaboration, Prusti, T., de Bruijne, J.~H.~J., et al.\ 2016, \aap, 595, A1. doi:10.1051/0004-6361/201629272

\bibitem[Gaia Collaboration et al.(2020)]{2020arXiv201201533G} Gaia Collaboration, Brown, A.~G.~A., Vallenari, A., et al.\ 2020, arXiv:2012.01533

\bibitem[Huang et al.(2015)]{2015RAA....15.1240H} Huang, Y., Liu, X.-W., Zhang, H.-W., et al.\ 2015, Research in Astronomy and Astrophysics, 15, 1240. doi:10.1088/1674-4527/15/8/010

\bibitem[Huang et al.(2020)]{2020arXiv201107172H} Huang, Y., Yuan, H., Li, C., et al.\ 2020, arXiv:2011.07172

\bibitem[Kingma \& Ba(2014)]{2014arXiv1412.6980K} Kingma, D.~P. \& Ba, J.\ 2014, arXiv:1412.6980

\bibitem[López-Sanjuan et al.(2020)]{}  López-Sanjuan, C., Yuan, H.-B., Vázquez Ramió, H., et al.\ 2020, A\&A, submitted

\bibitem[Luo et al.(2015)]{2015RAA....15.1095L} Luo, A.-L., Zhao, Y.-H., Zhao, G., et al.\ 2015, Research in Astronomy and Astrophysics, 15, 1095. doi:10.1088/1674-4527/15/8/002

\bibitem[Ma{\'\i}z Apell{\'a}niz \& Weiler(2018)]{2018A&A...619A.180M} Ma{\'\i}z Apell{\'a}niz, J. \& Weiler, M.\ 2018, \aap, 619, A180. doi:10.1051/0004-6361/201834051

\bibitem[Niu et al.(2021)]{} Niu, Z.X., Yuan, H.-B., \& Liu, J.-F. 2021, \apj, re-submitted 

\bibitem[Onken et al.(2019)]{2019PASA...36...33O} Onken, C.~A., Wolf, C., Bessell, M.~S., et al.\ 2019, \pasa, 36, e033. doi:10.1017/pasa.2019.27

\bibitem[Pancino et al.(2012)]{2012MNRAS.426.1767P} Pancino, E., Altavilla, G., Marinoni, S., et al.\ 2012, \mnras, 426, 1767. doi:10.1111/j.1365-2966.2012.21766.x

\bibitem[Riello et al.(2020)]{2020arXiv201201916R} Riello, M., De Angeli, F., Evans, D.~W., et al.\ 2020, arXiv:2012.01916

\bibitem[Sandage \& Smith(1963)]{1963ApJ...137.1057S} Sandage, A. \& Smith, L.~L.\ 1963, \apj, 137, 1057. doi:10.1086/147584

\bibitem[Schlafly \& Finkbeiner(2011)]{2011ApJ...737..103S} Schlafly, E.~F. \& Finkbeiner, D.~P.\ 2011, \apj, 737, 103. doi:10.1088/0004-637X/737/2/103

\bibitem[Schlegel et al.(1998)]{1998ApJ...500..525S} Schlegel, D.~J., Finkbeiner, D.~P., \& Davis, M.\ 1998, \apj, 500, 525. doi:10.1086/305772

\bibitem[Sun et al.(2021)]{}  Sun, Y., Yuan, H.-B., et al.\ 2021, \apjs, to be submitted

\bibitem[Weiler(2018)]{2018A&A...617A.138W} Weiler, M.\ 2018, \aap, 617, A138. doi:10.1051/0004-6361/201833462

\bibitem[Yuan et al.(2015)]{2015ApJ...799..133Y} Yuan, H.-B., Liu, X.-W., Xiang, M.-S., et al.\ 2015a, \apj, 799, 133. doi:10.1088/0004-637X/799/2/133

\bibitem[Yuan et al.(2015)]{2015ApJ...799..134Y} Yuan, H.-B., Liu, X.-W., Xiang, M.-S., et al.\ 2015b, \apj, 799, 134. doi:10.1088/0004-637X/799/2/134






\end{thebibliography}

\end{document}